\documentclass{aastex}

\usepackage{natbib}
\usepackage{graphicx}
\usepackage{float}
\usepackage{spr-astr-addons}
\usepackage{url}

\newcommand{\NGC}{NGC\,7793}
\newcommand{\NGCt}{NGC\,300}
\newcommand{\Hii}{H\textsc{ii}}
\newcommand{\hii}{h\textsc{ii}}
\newcommand{\hiib}{h\textsc{ii}}

\newcommand{\SNR}{SNR}
\newcommand{\snr}{snr}
\newcommand{\snrb}{snr}

\newcommand{\BCG}{BCKG}
\newcommand{\bcg}{bckg}
\newcommand{\bcgb}{bckg}

\newcommand{\osnr}{optically selected SNR}
\newcommand{\rsnr}{radio selected SNR}

\newcommand{\hodge}{H69}

\begin{document}
\maketitle
\title{Radio-Continuum Study of the Nearby Sculptor Group Galaxies. Part 3: NGC 7793 at $\lambda$=12.2, 6 and 3 cm}
\author{
Timothy J. Galvin, Miroslav D. Filipovi\'{c}, Nicholas F. H. Tothill, Evan J. Crawford, Andrew N. O'Brien
}
\affil{University of Western Sydney, Locked Bag 1797, Penrith, NSW, 2751, Australia} 
\and
\author{Nicholas Seymour}
\affil{CSIRO Astronomy and Space Science, PO Box 76, Epping, NSW 1710, Australia}
\and
\author{Thomas G. Pannuti, Alekzander R. Kosakowski, Biswas Sharma}
\affil{Space Science Center, Department of Earth and Space Sciences, Morehead State University, 235 Martindale Drive, Morehead, KY 40351 USA}
\date{\today}

\begin{abstract}
We re-examine a series of archived centimetre radio-continuum observations ($\lambda$=16, 6 and 3 cm) focusing on \NGC\ using the Australia Telescope Compact Array. These new images are both very sensitive ($\sigma$=0.011 mJy/beam) and feature reasonably high angular resolution (down to \textless3\arcsec). Using these images, a total of 76 discrete radio sources are identified, of which 57 have been classified. We also studied the radio component of the micro-quasar NGC7793-S26 which shows two distinct regions of somewhat steep spectral index ($\alpha$) between --0.3 and --0.7. 
\end{abstract}


\section{Introduction}
As part of the Sculptor group galaxies, \NGC\ is at an approximate distance of 3.91~Mpc \citep{2003A&A...404...93K} and has been intimately examined throughout a number of multifrequency studies.  Of particular interest is the source NGC7793-S26, initially discovered by \citet{1997ApJS..108..261B}. Subsequent X-ray and Very Large Array (VLA) radio observations by \citet{1999A&A...341....8R} and \citet{2002ApJ...565..966P} note extended emission associated with a bright point source, designated S26. Further studies by \citet{2008AIPC.1010..303P}, \citet{2010Natur.466..209P},  \citet{2011AJ....142...20P}, and \citet{2012MNRAS.427..956D} examine this source in greater detail and suggest that it is micro-quasar in nature.

The Australia Telescope Compact Array (ATCA), having been upgraded with the Compact Array Broadband Backend (CABB) \citep{2011MNRAS.416..832W}, was used by \citet{2010atnf.prop.3023S} to resolve S26's radio lobe structure and map its spectral index. As a part of our study of nearby Sculptor Group galaxies, we re-examine this data with an additional focus of discrete sources within \NGC's field.  

Part 1 of this series of papers \citep{2012Ap&SS.340..133G} created a new set of highly sensitive and highly  resolved mosaics of \NGCt\ -- another Sculptor Group galaxy -- at $\lambda$ = 20~cm by combining ATCA and VLA data. Part 2, by \citet{2013Ap&SS.347..159O}, re-examines ATCA and VLA data of the NGC\,55 -- yet another Sculptor Group galaxy -- field at $\lambda$ = 20, 13, 6 and 3~cm. In this paper we re-examine archived centimetre radio-continuum observations ($\lambda$=12.2, 6 and 3~cm) focusing on \NGC\ using the ATCA with an additional focus on discrete sources. In \S 2 we describe the observational data and reduction techniques. We present and discuss our results in \S 3 and conclusion in \S 4.

\section{Data and data reduction}
 \subsection{Observational Data}

The data used in this study was obtained from the Australia Telescope Online Archive (ATOA)\footnote{http://atoa.atnf.csiro.au/}. Project C2096 was conducted over 5 non-consecutive days using the ATCA and CABB, which provided a 2~GHz spectral window for each of the observations. The observations used in this study are summarised in Table~\ref{table:observations}. 

In project C2096 we used PKS1934-638 and PKS2357-318 as its primary (flux and bandpass) and secondary (phase) calibrators, respectively. All observations used in this study were centred at RA (J2000)= 23$^h$57$^m$59.94$^s$ and DEC (J2000)=--32$^\circ$33\arcmin30.8\arcsec.

\begin{table} [h!]
\caption{ATCA CABB C2096 observations used in this study. The bandwidth of all observations is 2~GHz.}
\label{table:observations}
\begin{tabular}{c c c c}
\hline
Observation & Array & Central  & Time  \\
Date &  & Frequency & on source \\
& & (GHz) & (hours) \\
\hline
\phantom{ }6/08/2009 & 6D & 5.5, 9 & \phantom{ } 6.7 \\
\phantom{ }7/08/2009 & 6D & 5.5, 9 & \phantom{ } 6.2 \\
27/06/2010 & 6C & 5.5, 9 & 10.2 \\
28/06/2010 & 6C & 5.5, 9 & \phantom{ } 9.8 \\
29/06/2010 & 6C & 2.45 & \phantom{ } 9.9 \\
\hline
\end{tabular}
\end{table}

%Frequency Configuration 1
%  Channels  Freq(chan=1)  Increment  Restfreq     IFChain
%    2049       3.47400   -0.0010000   0.00000 GHz     1
%    1.425GHz =  21cm
%    3.474GHz = 8.6cm
%    
%
%Frequency Configuration 1
%  Channels  Freq(chan=1)  Increment  Restfreq     IFChain
%    2049       4.47600    0.0010000   0.00000 GHz     1
%    4.476GHz = 6.7cm
%    6.525GHz = 4.59cm

\subsection{Image Creation}

The \textsc{miriad} \citep{2011ascl.soft06007S} and \textsc{karma} \citep{2011ascl.soft02018G} software packages were used for data reduction and analysis. Typical \textsc{miriad} calibration and flagging procedures were then carried out, including the use of the guided automatic flagging task \textsc{pgflag}. 

Each observation was then imaged in order to verify its integrity. The task \textsc{invert} was used to produce a dirty map and beam. We found that supplying a Briggs robust parameter of 0.5 was the most effective in reducing the areas of sidelobe distortion. As we imaged the full spectral window provided by CABB, \textsc{mfclean} \citep{1994A&AS..108..585S} was used to deconvolve the multi-frequency synthesised dirty map. Finally, \textsc{restor} and \textsc{linmos} were used to convolve and correct for primary beam attenuation. 

Once we verified that sufficient data flagging and calibration had been carried out for each individual observation, a single image was created by combining datasets from all epochs for each particular frequency. 
 
%While visually inspecting the images we note that there were a number of sources outside the central field of each map that suffered from side-lobe distortion. \textsc{mfclean} requires the beam to be three times the size of the area to clean. For this reason we invoked the `double' option in \textsc{invert} when producing the dirty maps and beam files, then set the \textsc{mfclean} region to `quarter' when cleaning the image. Consequently, although the central field of the image was cleaned appropriately, the outer field of each map was not. 

%The hour angle coverage for the $\lambda$=2.4GHz image was as complete as both the $\lambda$=5.5 and 3 cm images. However, both $\lambda$=5.5 and 9.0GHz had more overlap in its coverage. This can be attributed to the more hours of time on source across more observing days for these two frequencies. The $uv$ coverage for each image was also fairly complete as the full 2 GHz observing window was used to produce each image. For this reason, the 6th antenna was not excluded in our imaging.

\section{Results and Discussion}

\begin{figure*}
\caption{Project C2096 at $\lambda$=12.2 cm in mJy/beam. The red ellipse highlights the area that was examined for sources. The beam is 7.6$\arcsec\times$3.8\arcsec, as represented by the blue ellipse in the bottom left corner.  The primary beam was blanked after 23\arcmin.}
\label{fig:ngc77932450}
\includegraphics[angle=-90,scale=0.6,trim=4.5cm 2.8cm 3cm 2cm, width=\textwidth, clip=true]{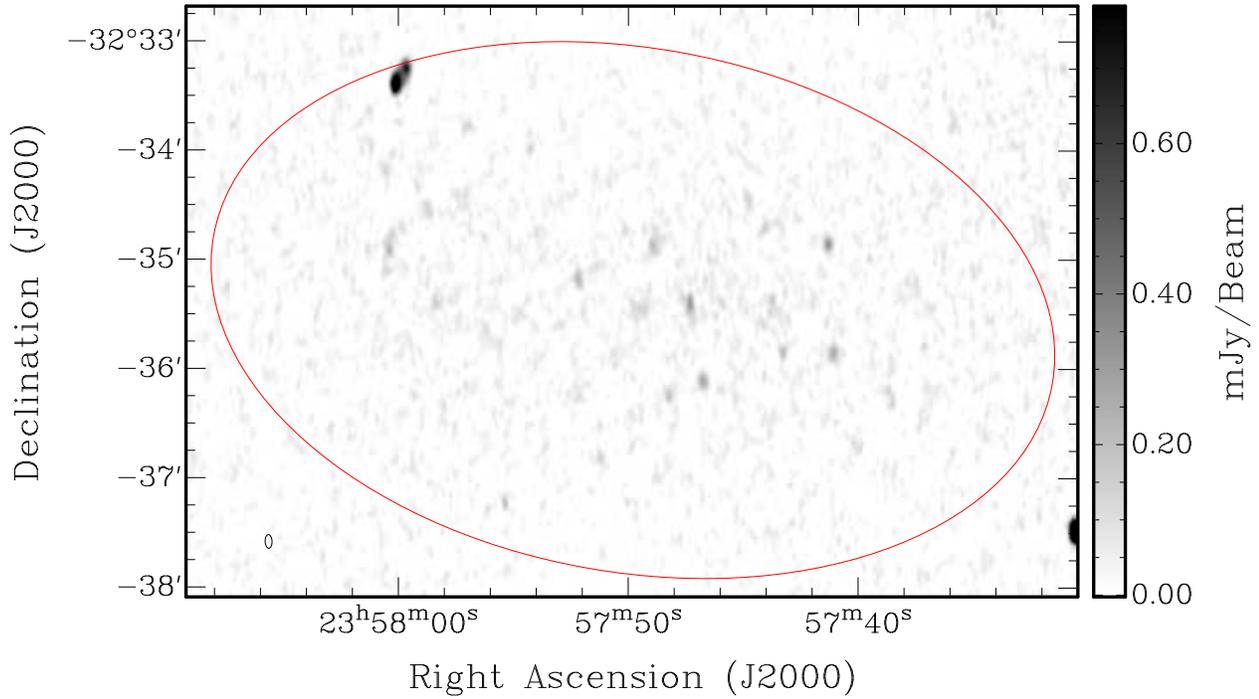}
\end{figure*}

\begin{figure*}
\caption{Project C2096 at $\lambda$=6 cm in mJy/beam. The red ellipse highlights the area that was examined for sources. The beam is 3.6$\arcsec\times$1.8\arcsec, as represented by the blue ellipse in the bottom left corner. The primary beam was blanked after 9.08\arcmin.}
\label{fig:ngc77935500}
\includegraphics[angle=-90,scale=0.6,trim=4cm 1.5cm 2.5cm 1.2cm,width=\textwidth,clip=true]{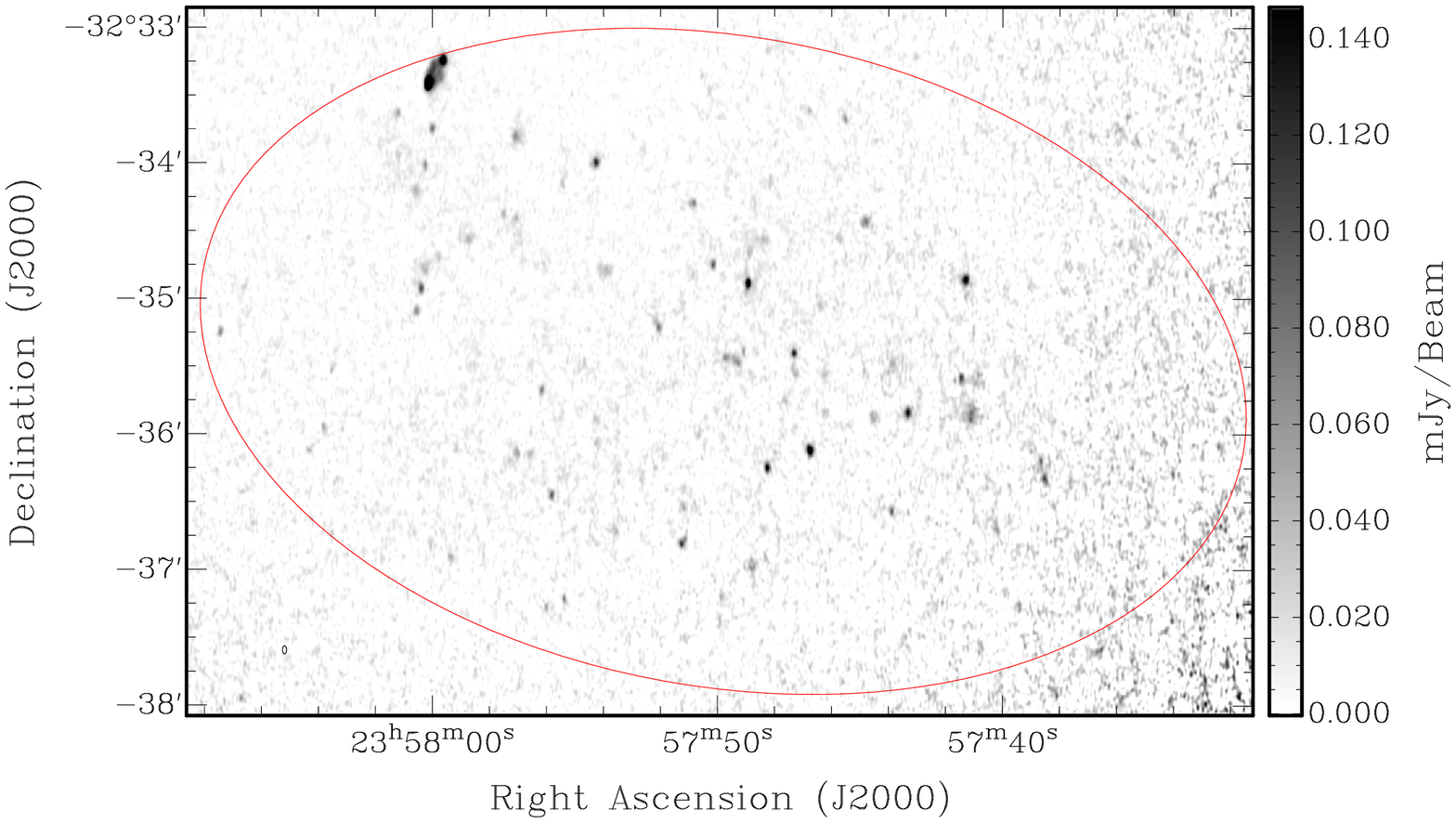}
\end{figure*}

\begin{figure*}
\caption{Project C2096 at $\lambda$=3 cm in mJy/beam. The red ellipse highlights the area that was examined for sources. The beam is 2.2$\arcsec\times$1.1\arcsec, as represented by the blue ellipse in the bottom left corner. The primary beam was blanked after 5.86\arcmin.}
\label{fig:ngc77939000}
\includegraphics[angle=-90,scale=0.6,trim=4cm 1.5cm 2.5cm 1cm,width=\textwidth, clip=true]{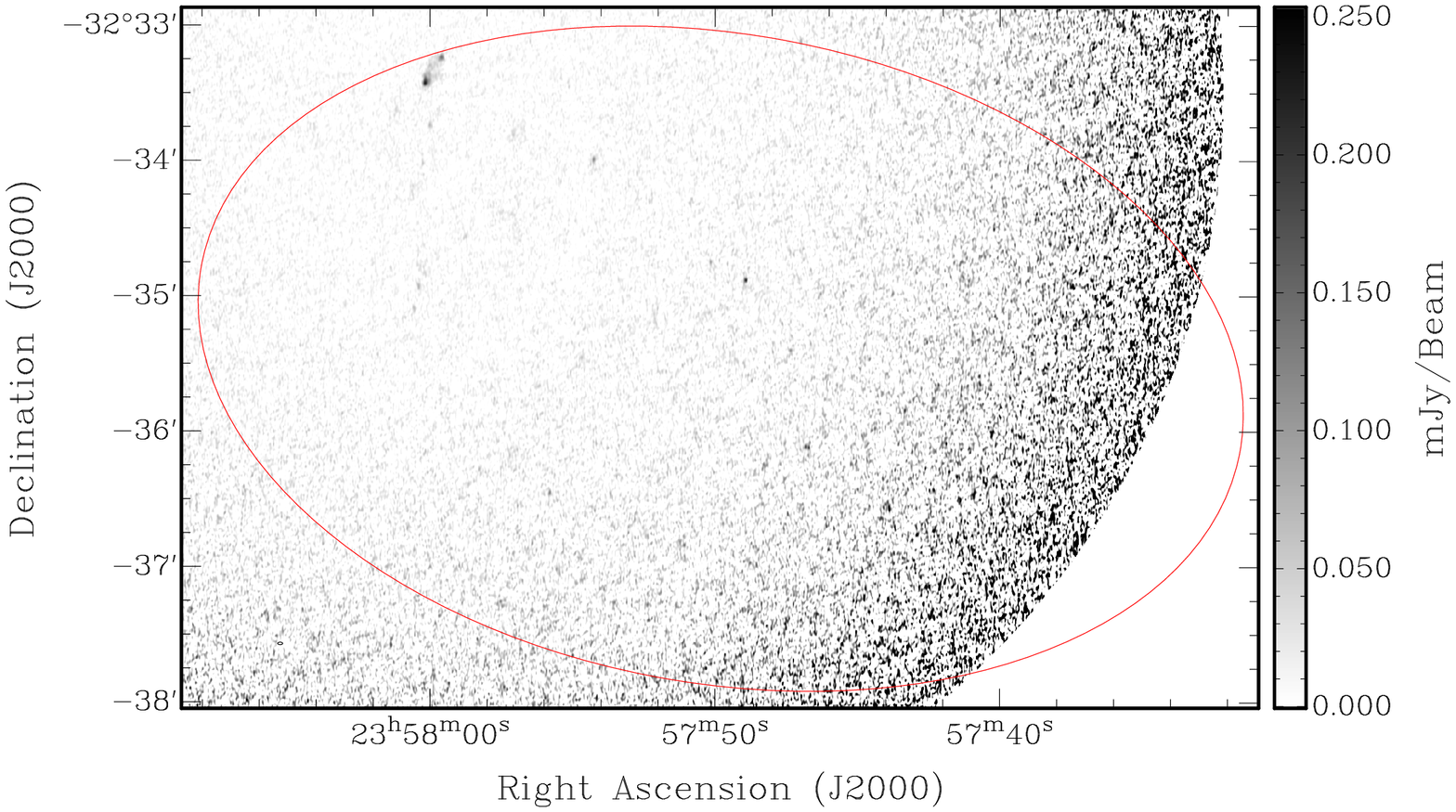}
\end{figure*}

The final images produced in this study are presented in Figs.~\ref{fig:ngc77932450}, \ref{fig:ngc77935500} and \ref{fig:ngc77939000}. Each image was produced using the collective dataset for that particular frequency. We summarise the details and parameters of each image in Table~\ref{table:images}.

Upon inspecting the final images we find that $\lambda$=12.2~cm (Fig.~\ref{fig:ngc77932450}) had a significantly higher RMS noise level (1$\sigma$=0.30 mJy/beam) than $\lambda$=6 and 3~cm  (1$\sigma$=0.011 and 0.015 mJy/beam in Figs.~\ref{fig:ngc77935500} and \ref{fig:ngc77939000} respectively). This was expected as \NGC\ at $\lambda$=12.2~cm was only observed for $\sim$10 hours, where as $\lambda$ = 6/3 cm was observed for $\sim$33 hours. As noise increased towards the edge of the primary beam, the adopted RMS values represent the upper limit in terms of local noise for each image. In the case of Fig.~\ref{fig:ngc77939000}, approximately one third of the field of interest was ignored due to the increased confusion at the edge of the primary beam.

\begin{table}[h]
\caption[l]{The image details of ATCA project C2096.}
\label{table:images}
\begin{tabular}{c c c c c}
\hline 
$\nu$ & Beam Size & PA & RMS ($\sigma$) & Figure\\
(GHz) & (arcsec) & $^\circ$ & (mJy/beam) & \# \\
\hline
2.4 & 7.6$\times$3.8 & 2.4 & 0.030 & \ref{fig:ngc77932450} \\
5.5 & 3.6$\times$1.8 & 1.0 & 0.011 & \ref{fig:ngc77935500} \\
9.0 & 2.2$\times$1.1 & -1.1& 0.015 & \ref{fig:ngc77939000} \\
\hline
\end{tabular}
\end{table}

 \subsection{Discrete sources within the field of NGC7793}

\begin{figure*}
\caption{Spectral index pixel map of \NGC\  as calculated from $\lambda$= 12.2, 6 and 3 cm data. The image is in terms of spectral index $\alpha$, where $\alpha$ is defined as $S \propto \nu^{\alpha}$. The red ellipse highlights the area that was examined for sources and the colour bar on the right reflects the spectral index value. The synthesised beam, represented by the blue ellipse in the lower left hand corner, is $7.56\arcsec$ x  $3.80\arcsec$.}
\label{fig:ngc7793-specmap}
\center\includegraphics[angle=-90,scale=0.4,trim=5cm 3cm 3.5cm 2cm, width=\textwidth,clip=true]{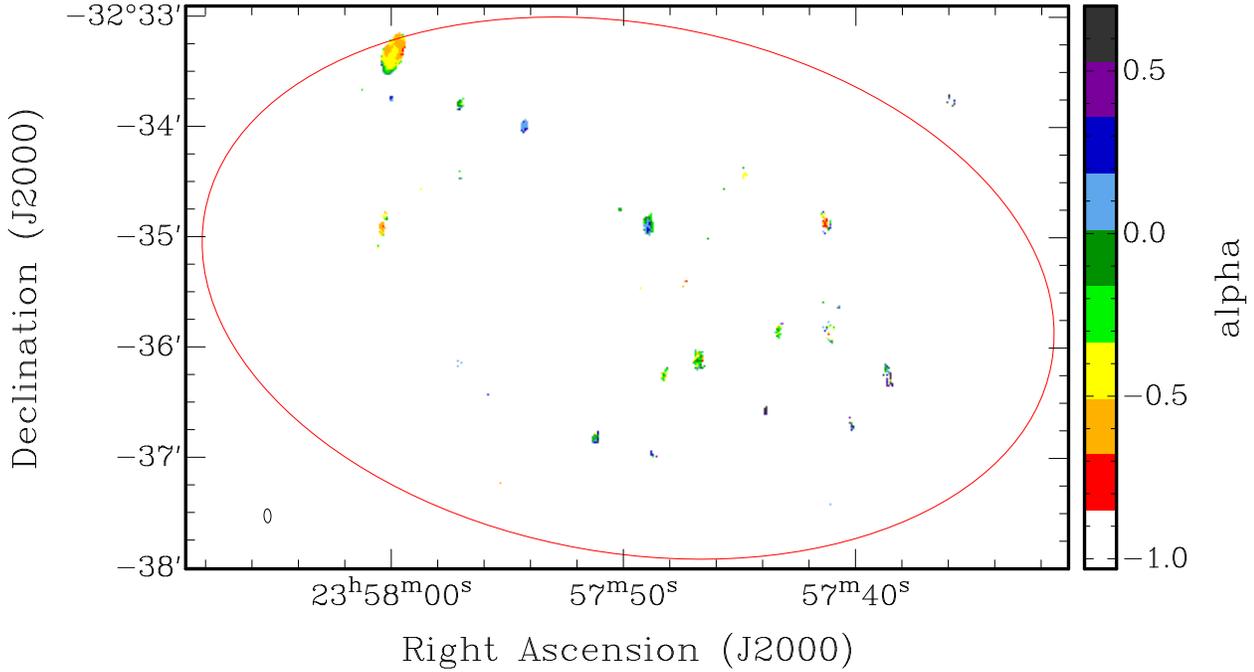}
\end{figure*}

We initially searched for sources in each image using the automated \textsc{miriad} task \textsc{sfind}. Using the returned results as a starting point list, we then reviewed each listed source in the field to evaluate its reliability. A number of proposed sources from \textsc{sfind} were discounted as they were either artefacts of side-lobes around particularly strong points or were outside the field of \NGC. After removing such sources, we then manually examined each image to identify any remaining sources which were not found by \textsc{sfind}. The red ellipse in Figs.~\ref{fig:ngc77932450}, \ref{fig:ngc77935500}, \ref{fig:ngc77939000} and \ref{fig:ngc7793-specmap} highlights the field which was searched for sources and was constructed to tightly cover the centre of \NGC\ as seen from the Digital Sky Survey \footnote{\url{http://archive.stsci.edu/cgi-bin/dss_form}} (DSS) optical image. The size of the major and minor axis of this ellipse is approximately 7.78\arcsec\ by 4.90\arcsec at a position angle of 280$^\circ$.

A total of 76 discrete sources above 3$\sigma$ (0.090, 0.033 and 0.045 mJy/beam at wavelengths 12.2, 6 and 3 cm (Table~\ref{table:sources2}; Column 5, 6 and 7 respectively) were identified within the field of \NGC. Using a 5$\arcsec$ search radius, a value adopt because of the beam size of Fig.~\ref{fig:ngc77932450}, a total of 39 sources were found at more than one wavelength. In such cases we estimated sources spectral index, $\alpha$, defined as $S \propto \nu^{\alpha}$ (Table~\ref{table:sources2}; Column 12). When calculating this spectral index value we assumed a error in flux density of 10\% for each measurement. Position, integrated flux densities and spectral indices of these catalogued sources are presented in Table~\ref{table:sources2}. The positions listed in Table~\ref{table:sources2} refer to the location of each source from the highest frequency image from which it was found. Due to the smaller primary beam at $\lambda$ = 3cm (Fig.~\ref{fig:ngc77939000}), approximately one third of the search ellipse is too noisy for source detection. The 28 sources within this flagged area have been marked in Table.~\ref{table:sources2} with $\dagger$. 

%ue to the smaller size of the primary beam for Fig.~\ref{fig:ngc77939000}, approximately one third of the field of interest was considered unreliable and was not searched.

\begin{figure*}
\caption{A histogram showing the distribution of flux densities of sources detected at $\lambda$ = 12.2, 6 and 3 cm (Table. \ref{table:sources2}; Column 5, 6 and 7). All sources have been binned in 0.05 mJy increments. Due to its significant flux density when compared to other discrete sources in this study, NGC7793-S26 (Table.~\ref{table:sources2}; Index 61) is excluded from this graph. The dotted line represents the 3$\sigma$ level.}
\label{fig:12fluxdist}
\center\input{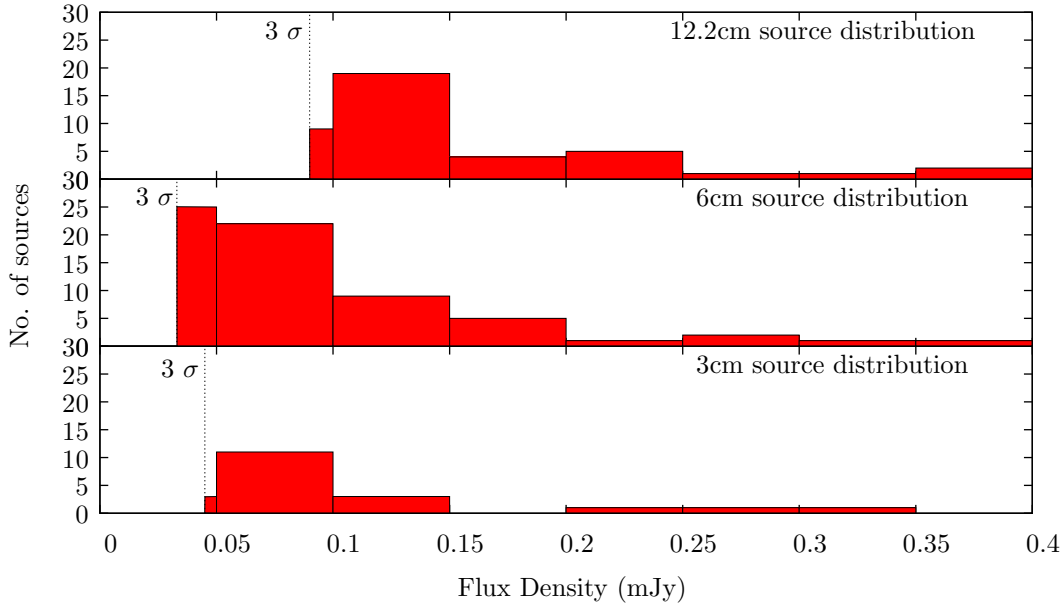}
\end{figure*}

\begin{figure*}
\caption{A histogram showing both the distribution of spectral indexes and classification of sources presented in Table. \ref{table:sources2}; Columns 12 and 13. J235743-323634 (Table.~\ref{table:sources2}; Index. 13) is excluded from this graph.}
\label{fig:specindex}
\center\includegraphics[angle=-90,scale=0.50,trim=0cm 0cm 0cm 0cm,clip=true]{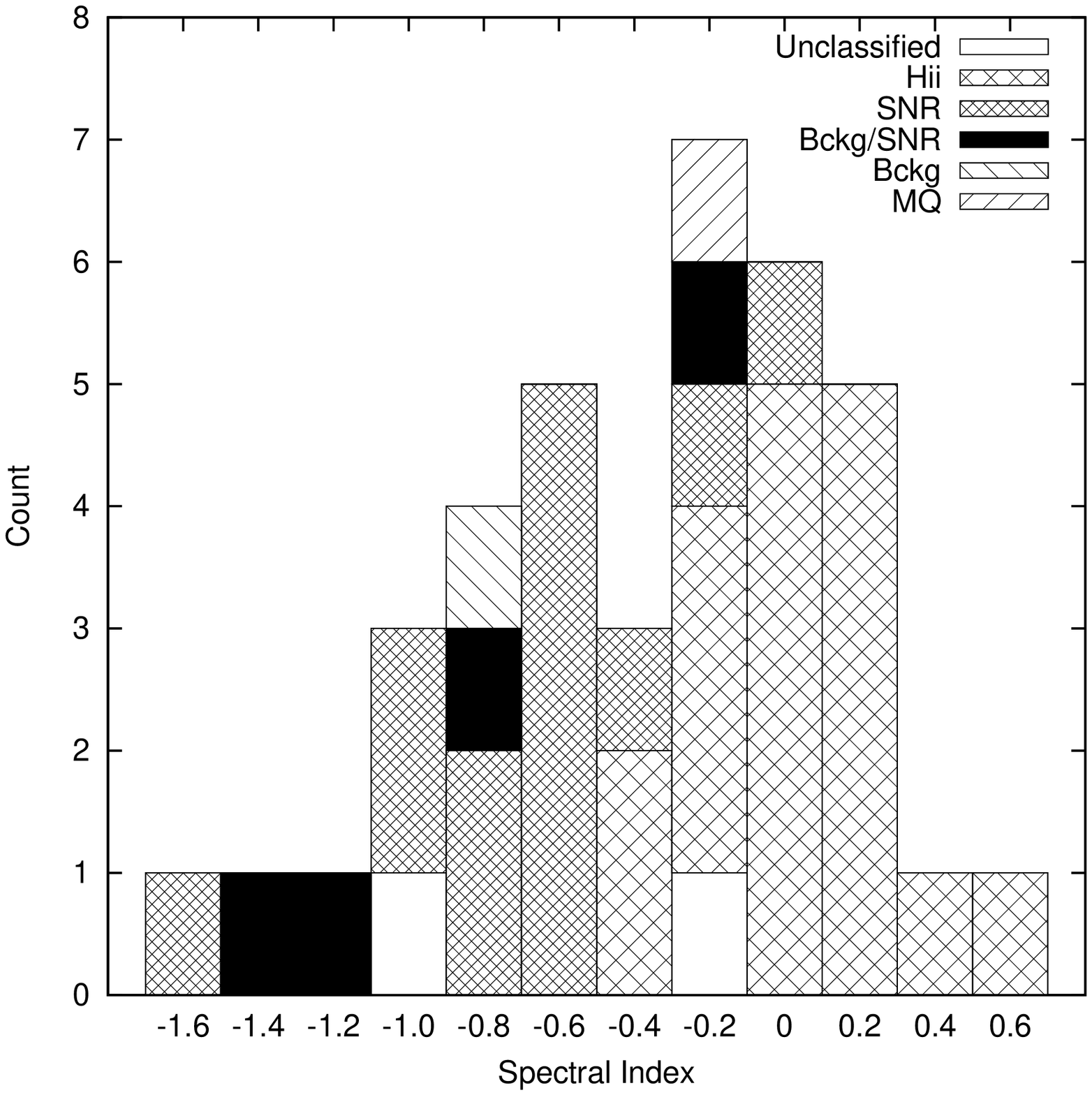}
\end{figure*}

Our new catalogue of 76 discrete radio-continuum sources in \NGC\ was then compared to the Infrared \citep{2003AAS...203.9010K}, Ultra-violet \citep{2009ApJ...703..517D}, Optical (DSS) and X-ray \citep{2011AJ....142...20P} catalogues and any source coincidences that were found using a search radius of 5\arcsec\ were noted in Table~\ref{table:sources2} (Columns 8, 9, 10 and 11 respectively).

%\begin{figure*}
%\caption{A histogram showing the distribution of flux densities of sources detected at $\lambda$ = 12.2, 6 and 3 cm (Table. \ref{table:sources2}; Column 5, 6 and 7). All sources have been binned in 0.05 mJy increments. NGC7793-S26 is excluded from this graph. The dotted line represents the 3$\sigma$ level.}
%\label{fig:12fluxdist}
%\includegraphics[angle=-90,scale=0.60,trim=5cm 0cm 0cm 0cm,clip=true]{fluxdist-stacked.eps}
%\end{figure*}

In Fig.~\ref{fig:12fluxdist} we show the distribution of integrated flux densities of detected sources at $\lambda$ = 12.2, 6 and 3 cm as described in Table~\ref{table:sources2} (Columns 5, 6 and 7 (excluding NGC7793-S26 (Table.~\ref{table:sources2}; Index 61)). We note that the majority of sources (85\%) have an integrated flux density that is below 10$\sigma$ flux density levels. 
Fig.~\ref{fig:specindex} shows the distribution (in terms of spectral index (Table. \ref{table:sources2}; Column 12) and classification (Table. \ref{table:sources2}; Column 13)) of 38 out of the 76 sources sources identified in this study. J235743--323634 (Table.~\ref{table:sources2}; Index 13) is excluded from this graph. We note two distinctive source groups; one peaking at $\alpha$=--0.6 and second at $\alpha$=--0.2. Out of 39 sources with an estimated spectral index, 24 (61.5\%) can be classified as having a flat or inverted spectrum with an $\alpha$ between 0.7 and --0.5.  The remaining 15 sources (38.5\%) have a steeper $\alpha$ between --0.5 to --1.7. 

 \subsection{Source Classification}

Seventy-six (76) radio-continuum sources were detected in our study. These sources have been classified into three main groups: background sources, Supernova Remnants (SNRs) and \Hii\ regions. Combining previous work and our new radio-continuum maps we classify a total of 57 sources. From these 57 sources, 37 are most likely to be \Hii\ regions, 14 to be SNRs, 1 to be a micro-quasar (see \citet{2008AIPC.1010..303P}, \citet{2010Natur.466..209P},  \citet{2011AJ....142...20P} and \citet{2012MNRAS.427..956D}), 1 to be a background galaxy (or AGN) and 4 which could be either an SNR or a background galaxy (or AGN). The remaining 19 radio-continuum sources still have no classification. We list each sources classification, where appropriate, in Table \ref{table:sources2}; Column 13. Strong candidates are listed as either \Hii, \SNR, MQ\ (micro-quasar) or \BCG\ (background galaxy/AGN), while weaker candidates are listed as \hii, \snr\  and \bcg. Fig.~\ref{fig:specindex} shows the distribution of these source classifications. 

A source was classified as a strong candidate SNR if it had both a steep negative spectral index and an X-Ray counterpart. If a source only had one of these properties it was classified as a weak candidate snr or bckg/snr. Likewise, if a source had a flat spectral index and an Infrared counterpart it was classified as a \Hii\ region candidate. If it only had one such property, it was classified as a weak \Hii\ region candidate. Each source was also visually inspected to determine if there was any intrinsic structure that would help in its identification. Three sources (Table \ref{table:sources2}; Index \#26, \#39 and \#67) have been classified as being weak candidate snrs or bckg/snr while having a shallow spectral index. These sources were classified as either weak candidate \snr s or \snrb/\bcgb\ based on how many counterparts were seen at additional frequency bands, the lack of extended emission or structure of the source, and the uncertainty associated with borderline steep spectral index values.   

% J235746-323606 (Table \ref{Table:sources2}; Index \#26) was classified as an \snr\ as it had a counterpart at each frequency band compared against, as well as being classified as an \snr\ by \citet{2002ApJ...565..966P}. J235750-323444 (Table \ref{Table:sources2}; Index \#39) was classed as a weak candidate \snrb/\bcgb\ as it was a point source and only had a optical counterpart. J235800-323411 (Table \ref{Table:sources2}; Index \#67) had counterparts at 3 different frequency bands (Infra-red, Ultra-violet and Optical) while having a spectral index of -0.27 $\pm$ 0.25.  

%\begin{figure*}
%\caption{The positional offset of sources common to both our study (Table.~\ref{table:sources2}) and either \citet{2002ApJ...565..966P}, which lists radio and optically select SNR candidates, or \citet{2011AJ....142...20P}, which lists \textit{Chandra} X-ray detections. The average positional difference in $\Delta$ RA and $\Delta$ DEC is --0.13$\arcsec$ (with a SD of 1.19$\arcsec$) and --0.08$\arcsec$ (with a SD of 0.61$\arcsec$) respectively.}
%\label{fig:offset}
%\center\includegraphics[angle=-90,scale=0.40,trim=0cm 0cm 0cm 0.5cm,clip=true]{offset.eps}
%\end{figure*}

After classifying our sources we then compared them to two additional catalogues. The first catalogue, from \citet{2002ApJ...565..966P}, listed a total of 33 radio and optically selected SNR candidates. A total of 5 sources were found to be common between both studies, of which we had classified 4 as SNRs and 1 which was not classified. The second catalogue, from \citet{1969ApJS...18...73H}, listed \Hii\  regions. When compared to our own we find a total of 17 source matches to be common. Of these 17 sources, we had classified 12 as \Hii\  regions and 4 as \SNR s. The remaining source was the micro-quasar NGC7793-S26. \citet{1969ApJS...18...73H} used two photographs, a yellow image and a yellow subtracted H$_{\alpha}$ image, to identify \Hii\ regions. This explains why 5 sources were classified differently when compared to \citet{1969ApJS...18...73H}, as these sources would have similar optical emission.  These sources are noted in Table~\ref{table:sources2}.

If a source has an X-ray counterpart when compared with \citet{2011AJ....142...20P} (Table~\ref{table:sources2}; Column 11) or was listed as an SNR candidate in \citet{2002ApJ...565..966P} (as noted in Table~\ref{table:sources2}; Column 14), we compared the listed positions from both catalogues. We find that the average positional difference in $\Delta$ RA and $\Delta$ DEC is --0.13$\arcsec$ (with a SD of 1.19$\arcsec$) and --0.08$\arcsec$ (with a SD of 0.61$\arcsec$) respectively. %This positional comparison is presented in Fig.~\ref{fig:offset}. 

We find 57 sources in common between radio (Figs. \ref{fig:ngc77932450}, \ref{fig:ngc77935500} and \ref{fig:ngc77939000}) and at least one additional frequency band (Infrared, Ultra-violet, Optical and X-ray). Of these 5 frequency bands, there were 3 sources which were seen in all five, 38 in four, 11 in three and 3 in two.  

 \subsection{Spectral Map}

Fig. \ref{fig:ngc7793-specmap} shows a three point spectral map of NGC7793. Each pixel represents the spectral index as measured across Figs.~\ref{fig:ngc77932450}, \ref{fig:ngc77935500} and \ref{fig:ngc77939000} after convolving the images to the largest beam size (12.2~cm). Pixels below the 3 sigma RMS noise level for each image were ignored.

The compact sources within this study exhibit a shallow (flatish) spectral index of between $-0.2$ $\textgreater$ $\alpha$ $\textgreater$ $-0.4$. This indicates dominance of regions with high thermal radiation caused by dense star formation. We can also see occasional steeper areas where $\alpha$ $\textless$ $-0.4$, implying a dominance of non-thermal radiation possibly caused by either synchrotron or inverse-Compton radiative mechanisms. Objects which radiate using this mechanism include SNRs and energetic jets.

 \subsubsection{Micro-quasar S26}

\citet{2010Natur.466..209P} suggest that NGC 7793-S26 (Table \ref{table:sources2}; Index 61) is powered by a black hole expelling relativistic jets of matter. Through the interactions with the surrounding interstellar medium (ISM), two bubbles have formed either side of the central black hole.

\begin{figure}[!h]
\caption{Spectral index pixel map of NGC7793-S26 as calculated from $\lambda$= 12.2, 6 and 3 cm data. The image is in terms of spectral index $\alpha$,where $\alpha$ is defined as $S \propto \nu^{\alpha}$. The colour bar on the right reflects the spectral index value. The synthesised beam, represented by the blue ellipse in the lower left hand corner, is $7.56\arcsec$ $\times$  $3.80\arcsec$. The contour levels are 0.01 mJy to 0.80 mJy with 0.05 mJy increments and are representative of the 6 cm data (Fig. \ref{fig:ngc77935500}).}
\label{fig:S26-specmap}
\includegraphics[angle=-90,trim=0.5cm 2cm 0.5cm 3.1cm, width=0.5\textwidth,clip=true]{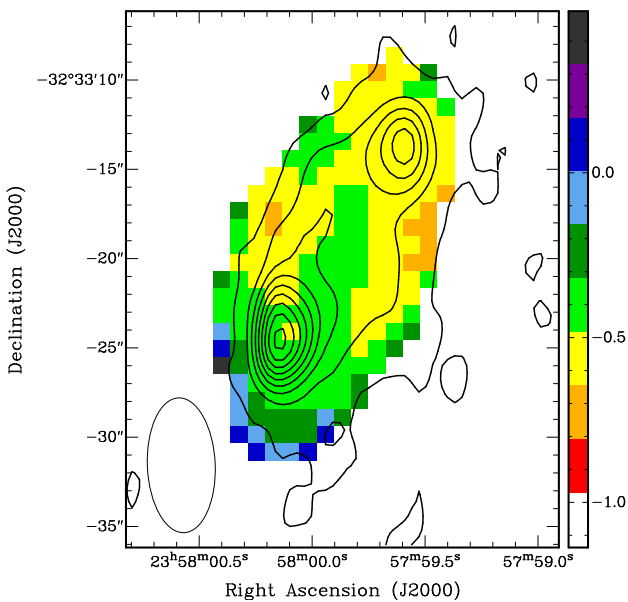}
\end{figure}

Fig.~\ref{fig:S26-specmap} shows NGC7793-S26 from the spectral map (as discussed above) overlaid with contours from Fig.~\ref{fig:ngc77935500} ($\lambda$=6 cm). When overlaid with contours, the two ISM bubbles can be clearly seen. The north most bubble {\bf(J2000 RA=23$^h$57$^m$59.7$^s$, DEC=-32$^\circ$33\arcmin13.00\arcsec)} has an average spectral index of approximately $\alpha\sim$--0.54 and range of --0.74 to --0.24, while the south most bubble {\bf(J2000 RA=23$^h$58$^m$00.1$^s$, DEC=-32$^\circ$33\arcmin25.00\arcsec)} has a slightly flatter average spectral index of $\alpha\sim$--0.43 with a range of --0.65 to --0.04. We also note a slightly positive index (approximately 0 $\textless$ $\alpha$ $\textless$ 0.2) on the outskirts of the southern lobe.

The peak flux intensity of the southern lobe is measured at 1.28~mJy/beam, 0.42~mJy/beam and 0.23~mJy/beam at 12.2, 6 and 3~cm respectively. The northern lobe is somewhat weaker with a peak flux intensity of 0.67~mJy/beam at 12.2~cm, 0.24~mJy/beam at 6~cm and 0.14~mJy/beam at 3~cm. 

%Fig. \ref{fig:s26-specmap} shows a spectr 

%Using the three calibrated frequencies, we created a spectral map of NGC7793. We first re-imaged each set of observations to force matching cell, FWHM and parallactic angle values. Each pixel in the spectral map represents the spectral index as calculated using the corresponding pixels from the three matching The final result Pixels that were below the noise level were ignored. 

\section{Conculsion}
Using archived centimetre radio-continuum observations ($\lambda$=12.2, 5.5 and 3~cm) of \NGC\ we catalogue a total of 76 discrete radio sources. Of these 76 sources, we classified a total of 37 sources as \Hii\ regions, 14 as \SNR s, 1 as a micro-quasar, 1 as a background galaxy and 4 as either background galaxies or \SNR s. A spectral index map of \NGC\ was created which exhibited a mostly shallow spectral index value, indicating dominance of regions with high thermal radiation caused by dense star formation. We also studied the radio component of the micro-quasar NGC7793-S26, and showed that on average it has a somewhat steep spectral index of between --0.3 and --0.7.

\section{Acknowledgements}

The Australia Telescope Compact Array is part of the Australia Telescope National Facility which is funded by the Commonwealth of Australia for operation as a National Facility managed by CSIRO.  This paper includes archived data obtained through the Australia Telescope Online Archive (\url{http://atoa.atnf.csiro.au}). The Digitized Sky Surveys were produced at the Space Telescope Science Institute under U.S. Government grant NAG W-2166. The images of these surveys are based on photographic data obtained using the Oschin Schmidt Telescope on Palomar Mountain and the UK Schmidt Telescope. The plates were processed into the present compressed digital form with the permission of these institutions. This research has made use of Aladin, SIMBAD and VizieR, operated at the CDS, Strasbourg, France. We used the {\sc karma} software package developed by the ATNF. 
Nicholas Seymour is the recipient of an ARC Future Fellowship.
We would also like to thank the anonymous referee whose feedback greatly improved the quality of this paper.

\begin{table*}
\caption{List of point sources in the \NGC\  field at $\lambda$ = 12.2, 6 and 3 cm. RA (3) and Dec (4) are in J2000 coordinates and are taken from the image with the highest resolution that a source was found in. Column 8 is the best fit spectral index for all flux measurements of a sources. {\bf When calculating this spectral index value we assumed a error in flux density of 10\% for each measurement.} Blank cells indicate no measurements were detect. The Infrared ID (8), Ultra-Violent ID (9), Optical ID (10) and X-ray ID (11) are from \citet{2003AAS...203.9010K}, \citet{2009ApJ...703..517D}, Digital Sky Survey and \citet{2011AJ....142...20P} respectively.  $\dagger$ represents a point that could not be distinguished from noise as it was outside the {\bf sensitive area of the} primary beam. Sources which are a strong candidate for their respective classification are presented in upper case, while less robust sources are presented in lower case. The notes `\osnr'\  and `\rsnr'\  denotes SNR candidates and their method of identification as listed by \citet{2002ApJ...565..966P}, while `\hodge'\ denotes sources classified as \Hii\ by \citet{1969ApJS...18...73H}.}
\label{table:sources2}
\scalebox{0.55}{
\begin{tabular}{cccccccccccccl}
\hline
  \multicolumn{1}{c}{1} &
  \multicolumn{1}{c}{2} &
  \multicolumn{1}{c}{3} &
  \multicolumn{1}{c}{4} &
  \multicolumn{1}{c}{5} &
  \multicolumn{1}{c}{6} &
  \multicolumn{1}{c}{7} &
  \multicolumn{1}{c}{8} &
  \multicolumn{1}{c}{9} &
  \multicolumn{1}{c}{10} &
  \multicolumn{1}{c}{11} &
  \multicolumn{1}{c}{12} &
  \multicolumn{1}{c}{13} &
  \multicolumn{1}{l}{14} \\

 \multicolumn{1}{c}{Index} &
 \multicolumn{1}{c}{Source} &
  \multicolumn{1}{c}{RA (J2000)} &
  \multicolumn{1}{c}{Dec (J2000)} &
  \multicolumn{1}{c}{$S_{16cm}$} &
  \multicolumn{1}{c}{$S_{6cm}$} &
  \multicolumn{1}{c}{$S_{3cm}$} &
  %Spitzer IRAC
  \multicolumn{1}{c}{IR} &
  %GALEX UV
  \multicolumn{1}{c}{UV} &
  %DSS
  \multicolumn{1}{c}{Optical} &
  %Tom Pannuti
  \multicolumn{1}{c}{X-ray} &
  \multicolumn{1}{c}{Spectral Index} &
  \multicolumn{1}{c}{Type} &
  \multicolumn{1}{l}{Notes} \\
  
   \multicolumn{1}{c}{} &
 \multicolumn{1}{c}{name} &
  \multicolumn{1}{c}{(h m s)} &
  \multicolumn{1}{c}{($^\circ$ \arcmin\space\arcsec)} &
  \multicolumn{1}{c}{(mJy)} &
  \multicolumn{1}{c}{(mJy)} &
  \multicolumn{1}{c}{(mJy)} &
  %Spitzer IRAC
  \multicolumn{1}{c}{ID} &
  %GALEX UV
  \multicolumn{1}{c}{ID} &
  %DSS
  \multicolumn{1}{c}{ID} &
  %Tome Pannuti
  \multicolumn{1}{c}{ID} &
  \multicolumn{1}{c}{$\alpha$} &
 \multicolumn{1}{l}{} &
  \\
\hline
     1 & J235737-323547 &  23:57:37.0 & --32:35:47.56 & 0.109 & & $\dagger$ & & & & &  & & \\
     2 & J235737-323512 &  23:57:37.5 & --32:35:12.80 & 0.100 & &$\dagger$ & & & & &  & & \\
     3 & J235737-323511 &  23:57:37.8 & --32:35:11.85 & 0.096 & &$\dagger$ & & & & &  & & \\
     4 & J235738-323619 &  23:57:38.5 & --32:36:19.31 & 0.120 & 0.120 &$\dagger$ & Y & Y & Y & & 0.00 $\pm$ 0.25 & \Hii &\hodge \\
\smallskip
     5 & J235738-323611 &  23:57:38.6 & --32:36:11.73 & 0.140 & 0.088 &$\dagger$ & Y & Y & Y & & -0.57 $\pm$ 0.25 & \snr & \hodge\\
     6 & J235739-323459 &  23:57:39.5 & --32:34:59.35 & 0.095 & &$\dagger$ & & & & &  & & \\
     7 & J235741-323552 &  23:57:41.0 & --32:35:52.80 & 0.244 & 0.270 &$\dagger$ & Y & Y & Y & & 0.13 $\pm$ 0.25 & \Hii & \hodge\\
     8 & J235741-323451 &  23:57:41.3 & --32:34:51.21 & 0.336 & 0.260 &$\dagger$ & Y & Y & Y & & -0.32 $\pm$ 0.25 & \hii & \hodge \\
     9 & J235741-323534 &  23:57:41.4 & --32:35:34.74 & 0.093 & 0.127 &$\dagger$ & Y & Y & Y & & 0.39 $\pm$ 0.25 & \Hii & \\
\smallskip
    10 & J235743-323550 &  23:57:43.3 & --32:35:50.20 & 0.214 & 0.180 &$\dagger$ & Y & Y & Y & & -0.21 $\pm$ 0.25 & \hii & \hodge \\
    11 & J235743-323521 &  23:57:43.7 & --32:35:21.70 & 0.160 & & $\dagger$& Y & & & &  & & \\
    12 & J235743-323528 &  23:57:43.8 & --32:35:28.47 & & 0.047 &$\dagger$ & & & & &  & & \\
    13 & J235743-323634 &  23:57:43.8 & --32:36:34.04 & & 0.098 & 0.316 & Y & Y & Y & Y & 2.34 $\pm$ 0.41 & & \\
    14 & J235743-323532 &  23:57:44.0 & --32:35:32.52 & & 0.043 &$\dagger$ & & & & &  & & \\
\smallskip
    15 & J235743-323442 &  23:57:44.0 & --32:34:42.14 & 0.144 & & $\dagger$& Y & & & &  & \hiib & \\
    16 & J235744-323553 &  23:57:44.5 & --32:35:53.03 & & 0.090 & $\dagger$& Y & Y & Y & &  & \hiib &\hodge \\
    17 & J235744-323425 &  23:57:44.8 & --32:34:25.36 & 0.140 & 0.150 &$\dagger$ & Y & Y & & & 0.09 $\pm$ 0.25 & \hii & \\
    18 & J235744-323531 &  23:57:44.8 & --32:35:31.33 & & 0.038 &$\dagger$ & Y & & Y & &  & & \\
    19 & J235745-323339 &  23:57:45.5 & --32:33:39.84 & & 0.058 & 0.065 & Y & Y & Y & & 0.23 $\pm$ 0.41 & \Hii & \\
\smallskip
    20 & J235745-323432 &  23:57:45.8 & --32:34:32.41 & & 0.034 &$\dagger$ & Y & Y & & &  & \hiib & \\
    21 & J235745-323524 &  23:57:45.9 & --32:35:24.66 & 0.117 & &$\dagger$ & & & Y & &  & & \\
    22 & J235746-323533 &  23:57:46.2 & --32:35:33.37 & & 0.038 &$\dagger$ & Y & Y & Y & &  & \hii & \\
    23 & J235746-323550 &  23:57:46.2 & --32:35:50.21 & & 0.039 &$\dagger$ & Y & Y & & &  & \hii & \\
    24 & J235746-323459 &  23:57:46.5 & --32:34:59.56 & & 0.034 & & Y & & Y & &  & \hii & \\
\smallskip
    25 & J235746-323511 &  23:57:46.6 & --32:35:11.15 & 0.104 & & & & & Y & &  & & \\
    26 & J235746-323606 &  23:57:46.7 & --32:36:06.98 & 0.269 & 0.370 & 0.270 & Y & Y & Y & Y & 0.04 $\pm$ 0.26 & \snr & \rsnr \\
    27 & J235746-323336 &  23:57:46.7 & --32:33:36.24 & & 0.043 & & Y & Y & Y & &  & \hii &  \\
    28 & J235747-323533 &  23:57:47.2 & --32:35:33.00 & & 0.042 & & Y & Y & Y & &  & \hiib & \\
    29 & J235747-323523 &  23:57:47.3 & --32:35:23.94 & 0.371 & 0.175 & & Y & & Y & Y & -0.93 $\pm$ 0.25 & \SNR & \osnr \\
\smallskip
    30 & J235748-323614 &  23:57:48.2 & --32:36:14.74 & 0.196 & 0.194 & 0.111 & Y & Y & Y & & -0.40 $\pm$ 0.30 & \snr & \rsnr ; \hodge \\
    31 & J235748-323654 &  23:57:48.3 & --32:36:54.78 & & 0.045 &$\dagger$ & & & & &  & & \osnr \\
    32 & J235748-323658 &  23:57:48.8 & --32:36:58.79 & & 0.061 &$\dagger$ & Y & Y & Y & &  & \hiib & \\
    33 & J235748-323452 &  23:57:48.9 & --32:34:52.72 & 0.210 & 0.310 & 0.214 & Y & Y & Y & & 0.06 $\pm$ 0.33 & \Hii & \\
    34 & J235749-323523 &  23:57:49.1 & --32:35:23.01 & & 0.054 & & Y & Y & Y & &  & \hii & \\
\smallskip
    35 & J235749-323528 &  23:57:49.2 & --32:35:28.23 & 0.126 & 0.065 & 0.053 & Y & Y & Y & & -0.68 $\pm$ 0.11 & \snrb& \\
    36 & J235749-323525 &  23:57:49.7 & --32:35:25.90 & 0.363 & 0.100 & & Y & Y & Y & & -1.59 $\pm$ 0.25 & \snrb & \\
    37 & J235749-323712 &  23:57:49.8 & --32:37:12.03 & & 0.048 &$\dagger$ & Y & Y & & &  & \hiib & \\
    38 & J235749-323723 &  23:57:49.9 & --32:37:23.65 & & 0.036 &$\dagger$ & & & & &  & & \\
    39 & J235750-323444 &  23:57:50.1 & --32:34:44.60 & 0.109 & 0.093 & 0.095 & & & Y & & -0.11 $\pm$ 0.06 & \bcgb/\snrb & \\
\smallskip
    40 & J235750-323600 &  23:57:50.6 & --32:36:00.72 & & 0.036 & & Y & Y & Y & &  & \hiib & \\
    41 & J235750-323417 &  23:57:50.8 & --32:34:17.33 & 0.091 & 0.096 & 0.061 & Y & Y & Y & & -0.27 $\pm$ 0.26 & \Hii & \hodge \\
    42 & J235751-323648 &  23:57:51.2 & --32:36:48.03 & 0.140 & 0.160 &$\dagger$ & Y & Y & Y & & 0.17 $\pm$ 0.25 & \hii & \\
    43 & J235752-323511 &  23:57:52.0 & --32:35:11.94 & 0.217 & 0.152 & 0.090 & Y & Y & Y & & -0.65 $\pm$ 0.17 & \snr & \hodge \\
    44 & J235752-323614 &  23:57:52.1 & --32:36:14.94 & 0.098 & 0.034 & & & & Y & & -1.31 $\pm$ 0.25 & \bcg/\snrb & \\
\smallskip
    45 & J235753-323446 &  23:57:54.0 & --32:34:46.49 & 0.097 & 0.150 & & Y & Y & Y & & 0.54 $\pm$ 0.25 & \hiib & \hodge \\
    46 & J235754-323359 &  23:57:54.2 & --32:33:59.10 & 0.132 & 0.203 & 0.114 & Y & Y & Y & & -0.05 $\pm$ 0.45 & \Hii & \hodge \\
    47 & J235754-323603 &  23:57:54.2 & --32:36:03.97 & & 0.038 & & Y & Y & Y & &  & \hiib & \\
    48 & J235754-323553 &  23:57:54.3 & --32:35:53.15 & & 0.034 & & Y & Y & Y & &  & \hiib & \\
    49 & J235755-323713 &  23:57:55.3 & --32:37:13.28 & 0.198 & 0.074 &$\dagger$ & & & & & -1.22 $\pm$ 0.25 & \bcg/\snrb& \\
\smallskip
    50 & J235755-323408 &  23:57:55.4 & --32:34:08.47 & & 0.034 & & & & & &  & & \\
    51 & J235755-323626 &  23:57:55.8 & --32:36:26.96 & & 0.110 & 0.122 & Y & & & & 0.21 $\pm$ 0.41 & \Hii & \\
    52 & J235755-323716 &  23:57:56.0 & --32:37:16.51 & & 0.056 &$\dagger$ & Y & Y & Y & &  & \hii & \\
    53 & J235756-323540 &  23:57:56.1 & --32:35:40.42 & 0.105 & 0.073 & 0.055 & Y & & Y & & -0.49 $\pm$ 0.03 & \Hii & \\
    54 & J235757-323607 &  23:57:57.0 & --32:36:07.97 & & 0.087 & & Y & Y & Y & &  & \hii & \hodge \\
\smallskip
    55 & J235757-323347 &  23:57:57.0 & --32:33:47.70 & 0.126 & 0.120 & 0.049 & Y & Y & Y & & -0.66 $\pm$ 0.47 & \snr & \hodge \\
    56 & J235757-323424 &  23:57:57.0 & --32:34:24.11 & & 0.044 & & Y & Y & Y & &  & \hii & \\
    57 & J235757-323422 &  23:57:57.5 & --32:34:22.26 & & 0.044 & & & & & &  & & \\
    58 & J235758-323523 &  23:57:58.3 & --32:35:23.88 & 0.156 & & & & & Y & &  & \snrb & \osnr \\
    59 & J235758-323433 &  23:57:58.7 & --32:34:33.58 & 0.130 & 0.085 & 0.051 & Y & Y & Y & & -0.70 $\pm$ 0.13 & \snr & \\
\smallskip
    60 & J235759-323654 &  23:57:59.3 & --32:36:54.52 & & 0.049 & 0.049 & Y & Y & & & 0.00 $\pm$ 0.41 & \hiib & \\
    61 & J235759-323317 &  23:58:00.0 & --32:33:17.20 & 109.000 & 94.300 & 74.210 & & & & & -0.28 $\pm$ 0.08 & MQ & NGC7793-S26; S26; Double source; \hodge \\
    62 & J235759-323343 &  23:58:00.0 & --32:33:43.80 & & 0.080 & 0.090 & Y & Y & Y & & 0.24 $\pm$ 0.41 & \hiib & \hodge \\
    63 & J235800-323447 &  23:58:00.2 & --32:34:47.31 & 0.140 & 0.060 & & Y & Y & Y & & -1.05 $\pm$ 0.25 & \snrb & \\
    64 & J235800-323359 &  23:58:00.3 & --32:33:59.79 & & 0.060 & 0.055 & Y & Y & Y & & -0.18 $\pm$ 0.41 & \hii & \\
\smallskip
    65 & J235800-323455 &  23:58:00.4 & --32:34:55.45 & 0.228 & 0.120 & 0.074 & Y & Y & Y & Y & -0.86 $\pm$ 0.05 & \SNR & \\
    66 & J235800-323505 &  23:58:00.6 & --32:35:05.21 & 0.119 & 0.067 & 0.063 & Y & Y & & & -0.51 $\pm$ 0.16 & \snr & \\
    67 & J235800-323411 &  23:58:00.6 & --32:34:11.74 & 0.131 & 0.105 & & Y & Y & Y & & -0.27 $\pm$ 0.25 & \snrb & \\
    68 & J235801-323337 &  23:58:01.2 & --32:33:37.02 & 0.094 & 0.075 & & Y & Y & Y & & -0.28 $\pm$ 0.25 & & \\
    69 & J235748-323433 &  23:57:48.4 & --32:34:33.27 & & 0.040 & & & Y & Y & &  & \hiib & \\
\smallskip
    70 & J235748-323439 &  23:57:48.9 & --32:34:39.30 & & 0.040 & & Y & Y & Y & &  & \hiib & \hodge \\
    71 & J235751-323632 &  23:57:51.4 & --32:36:32.79 & 0.094 & 0.050 & & & Y & Y & & -0.78 $\pm$ 0.25 &\bcg/\snrb & \\
    72 & J235753-323643 &  23:57:53.6 & --32:36:43.24 & & 0.043 & & Y & Y & Y & &  & \hiib & \hodge \\
    73 & J235807-323514 &  23:58:07.4 & --32:35:14.64 & 0.122 & 0.069 & 0.045 & & & & & -0.76 $\pm$ 0.04 & \bcg & \\
    74 & J235804-323607 &  23:58:04.3 & --32:36:07.19 & & 0.049 & & & & & &  & & \\
\smallskip
    75 & J235803-323556 &  23:58:03.8 & --32:35:56.19 & & 0.051 & & & & & &  & & \\
    76 & J235803-323531 &  23:58:03.5 & --32:35:31.52 & 0.096 & 0.041 & & & & & & -1.05 $\pm$ 0.25 & & \\
    
\hline\end{tabular}}\end{table*}

\bibliography{ngc7793-refs.v2}{}
\bibliographystyle{plainnat}
\end{document}